\documentclass[prl,notitlepage,
twocolumn,
longbibliography,superscriptaddress,amsmath,amssymb]{revtex4-2}

\usepackage{graphicx}  
\usepackage{dcolumn}   
\usepackage{bm}        
\usepackage{verbatim}   
\usepackage{bbold}
\usepackage{color}
\usepackage{xcolor}
\usepackage{amsthm}
\usepackage{amsmath}
\usepackage{ulem}
\usepackage{eufrak}

\usepackage{amsmath}
\usepackage{amsthm, amssymb}
\usepackage{slashed} 
\usepackage{graphicx}
\usepackage{xcolor}
\usepackage{bm}
\usepackage[colorlinks=true,citecolor=blue,linkcolor=magenta]{hyperref}
\usepackage{dsfont}
\usepackage{changepage}
\usepackage{array}

\def\a{\alpha}
\def\b{\beta}

\def\be{\begin{equation}}
\def\ee{\end{equation}}
\def\ba{\begin{align}}
\def\ea{\end{align}}
\def\>{\rangle}
\def\<{\langle}

\def\bea{\begin{eqnarray}}
\def\eea{\end{eqnarray}}
\def \T*{\text{T}^*}

\def \K3C60{\text{K}_3\text{C}_{60}}

\def \ETBr{\kappa\text{-(ET})_2\text{Cu}[\text{N(CN)}_2]\text{Br}}
\def \ETCl{\kappa\text{-(ET})_2\text{Cu}[\text{N(CN)}_2]\text{Cl}}
\def \ETSL{\kappa\text{-(ET})_2\text{Cu}_2\text{(CN)}_3}

\def\>{\rangle}
\def\<{\langle}

\theoremstyle{definition}

\theoremstyle{remark}

\begin{document}

\title{Superconducting-like response in driven systems near the Mott transition}

\author{Zhehao Dai}
\affiliation{
University of California, Berkeley, CA 94720, USA}
\author{Patrick A. Lee}
\affiliation{
Massachusetts Institute of Technology, Cambridge, MA 02139, USA
}
\date{\today}

\begin{abstract}
We point out that fractionalized bosonic charge excitations can explain the recently discovered photo-induced superconducting-like response in $\ETBr$, an organic metal close to the Mott transition. The pump laser exerts a periodic drive on the fractionalized field, creating a non-equilibrium condensate, which gives a Drude peak much narrower than the equilibrium scattering rate, hence superconducting-like response. Our proposal illuminates new possibilities of detecting fractionalization and can be readily tested in spin liquid candidates and in cold atom systems.
\end{abstract}

\maketitle

\begin{figure}[htb]
\begin{center}
\includegraphics[width=0.42\textwidth]{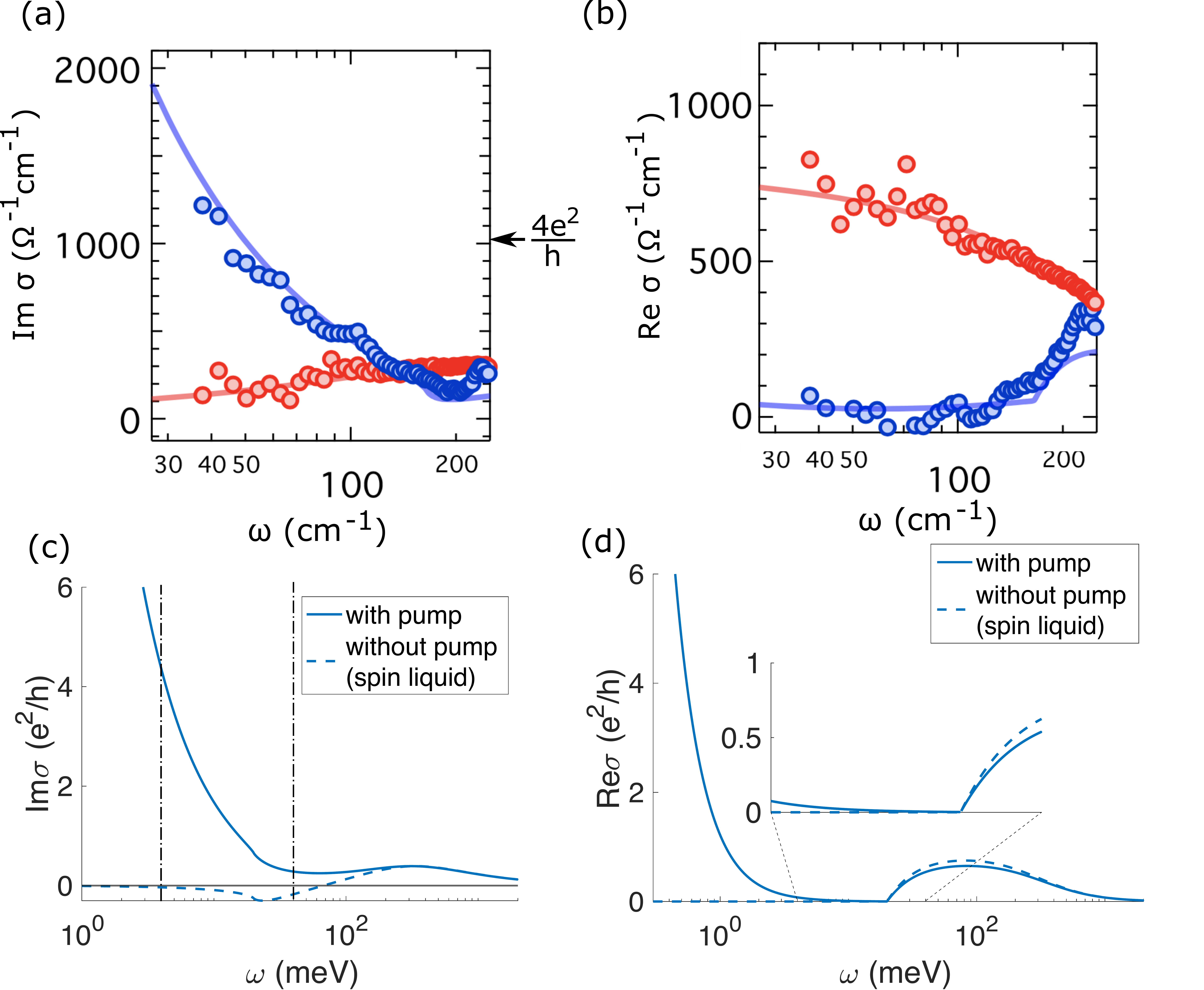}
\caption{(a-b) The measured real/imaginary part of the transient conductivity (blue dots) and the equilibrium conductivity (red dots) of the metallic sample $\ETBr$ at 30K (reproduced from Fig.3 in Ref.~\cite{PhysRevX.10.031028}). The solid blue line in (a) is constant times $1/\omega$. The frequency is shown in log scale. To facilitate comparison with the vertical scale of the theory curves, the black arrow on the right of (a) indicates a conductance of $4e^2/h$ per $\text{(ET)}_2$ layer. (c-d) Theoretical predictions of the real/ imaginary part of the conductivity (per layer, in units of $e^2/h$) of a driven spin liquid (solid line) and a spin liquid in equilibrium (dashed line), where we have used $\Delta=10$meV, $\rho_b = 3$meV, $\rho_f=40$meV, and $\tau^{-1}=1$meV. The dotted dashed lines in (c) and the inset of (d) represents the frequency range between 4meV and 40meV, comparable to the range shown in the experimental plot in (a-b). 
}
\label{Fig:conductivity}
\end{center}
\end{figure}

Ultrashort laser pulses have become powerful tools to perturb materials, to probe the underlying excitations, and to create new phases. Among the recent developments, photo-induced superconducting-like responses in various strongly correlated materials are particularly interesting and surprising. This phenomenon is found in multiple species of cuprate high-temperature superconductors~\cite{PhysRevB.89.184516,hu2014optically,cavalleri2018photo,PhysRevX.10.011053}, $\K3C60$~\cite{mitrano2016possible}, and $\ETBr$~\cite{PhysRevX.10.031028}. In all of the works cited above, a pump laser induces transient superconducting-like responses at temperatures much higher than the equilibrium superconducting transition temperature. A key experimental finding is a $1/\omega$ imaginary part of the conductivity after the pump (Fig.~\ref{Fig:conductivity}(a))~\cite{cavalleri2018photo,mitrano2016possible,PhysRevX.10.031028,PhysRevX.10.011053}, resembling the London equation of an equilibrium superconductor. In addition, $\ETBr$ and $\K3C60$ also show an emergent gap in the real part of the conductivity~\cite{,mitrano2016possible,PhysRevX.10.031028} (Fig.~\ref{Fig:conductivity}(b)). This has been interpreted as the photo-induced superconducting gap. Until now, it is still unclear what underlying mechanism produces the transient superconducting-like responses.

Almost all existing theoretical attempts assume some form of superconducting fluctuations. It is commonly assumed that the pump either creates a quasi-static superconducting phase ~\cite{cavalleri2018photo,mitrano2016possible,PhysRevX.10.031028} or dynamically excite collective modes which exist because of superconducting fluctuations.~\cite{PhysRevB.89.184516,hu2014optically,michael2020parametric} This assumption is reasonable for materials where superconducting fluctuations have been reported at relatively high temperatures. However, comparing the equilibrium properties with the transient response, we believe this scenario is unlikely to be the case for $\ETBr$. 

$\ETBr$ is a layered material with a nearly isotropic triangular lattice; the low-energy physics is believed to be described by a single-band Hubbard model with one electron per unit cell. The conducting band comes from electron orbitals of the ET molecular dimers. This material is well-known to be close to a Mott transition. In fact, $\ETCl$~\cite{PhysRevB.76.165113} and $\ETSL$~\cite{furukawa2018quasi}, which have ET layers with slightly different lattice parameters, are  Mott insulators. $\ETCl$ has an antiferromagnetic (AFM) order below 35K; $\ETSL$ is a well-studied candidate as a spin liquid with spinon Fermi surface. 
Being close to the insulating phase, $\ETBr$ behaves like an insulator in a broad temperature and frequency range, becoming metallic only at low temperature and frequency. The resistivity is very large at room temperature, rising with decreasing temperature and plunges rapidly around 60K~\cite{nam2013superconducting}. Indeed, at 50K the real part of the AC conductivity does not show a Drude peak at all, but rises with increasing frequency~\cite{PhysRevX.10.031028}. The standard interpretation is that a coherent Fermi liquid onsets only at $\T*\sim60$K~\cite{PhysRevX.10.031028, PhysRevB.76.165113}. The superconducting transition temperature is 12K in equilibrium, which is among the highest in the $\kappa\text{-ET}$ family. The superconducting gap has not been measured from the optical conductivity. However, after the pump, both the `transition temperature' and photo-induced 'superconducting-gap' are much larger than the equilibrium energy scale of superconductivity. The photo-induced superconducting-like response is observed up to 50K. At 15 to 30K, if we interpret half the optical gap (See Fig.~\ref{Fig:conductivity}(b)) as the induced `superconducting-gap', it is 11meV $\sim$ 130K, more than twice $\T*$~\cite{PhysRevX.10.031028}. It is hard to justify the existence of a quasi-static superconducting gap much larger than the coherent scale below which the electrons show metallic behavior. Some new point of view is called for. 



In a recent work, the authors propose a general mechanism for photo-induced superconducting-like response in materials with bosonic charge excitations~\cite{dai2021photoinduced}, where it is initially applied to preformed electron pairs related to quantum-fluctuating pair density waves in YBCO~\cite{PhysRevB.101.064502,dai2020exploring}. In this proposal, the laser pulse creates particle-hole excitations of the \textit{boson}; the density of the excitations grow exponentially at early time, inducing an AC conductivity just like ordinary superconductors, $\sigma(\omega)\propto i/\omega$, which we show is a consequence of Bose statistics. It may seem that the existence of bosonic charge excitation always implies pairing. Here we show that in materials close to the Mott transition, our formalism can be applied to fractionalized charge e bosons that emerges due to electron fractionalization.

The notion of fractionalization has been widely discussed in connection with the organic molecular materials consisting of ET dimers, especially in the spin liquid phases.
Starting from the Hubbard model at integer filling, Ref.~\cite{PhysRevLett.95.036403} established the slave boson (slave rotor) formalism, where the charge-e excitation has bosonic self statistics, which becomes physical in certain spin liquid phases. This is a highly nontrivial prediction yet to be confirmed in experiments. In this work we show that due to the emergent Bose statistics, the pump can still induce the exponentially growing excitations and produce a superconducting-like response, which we will clarify later. 
It follows that photo-induced superconducting-like response should appear in spin liquids which has no superconducting fluctuations at all! In particular, we propose to test this theory in the spin liquid candidates $\ETSL$ and $\beta'\text{-EtMe}_3\text{Sb[Pd(dmit}_2)]_2$. The requirement is that fractionalization occurs and the empty (holon) and doubly occupied  (doublon) states are charge excitations which obey Bose statistics at an intermediate energy given by the difference between the half the pump frequency and the Mott gap. An experimental observation of photo-induced superconducting-like response in these insulating materials would provide strong evidence for fractionalization at some intermediate energy scale.

\begin{figure}[t]
\begin{center}
\includegraphics[width=0.3\textwidth]{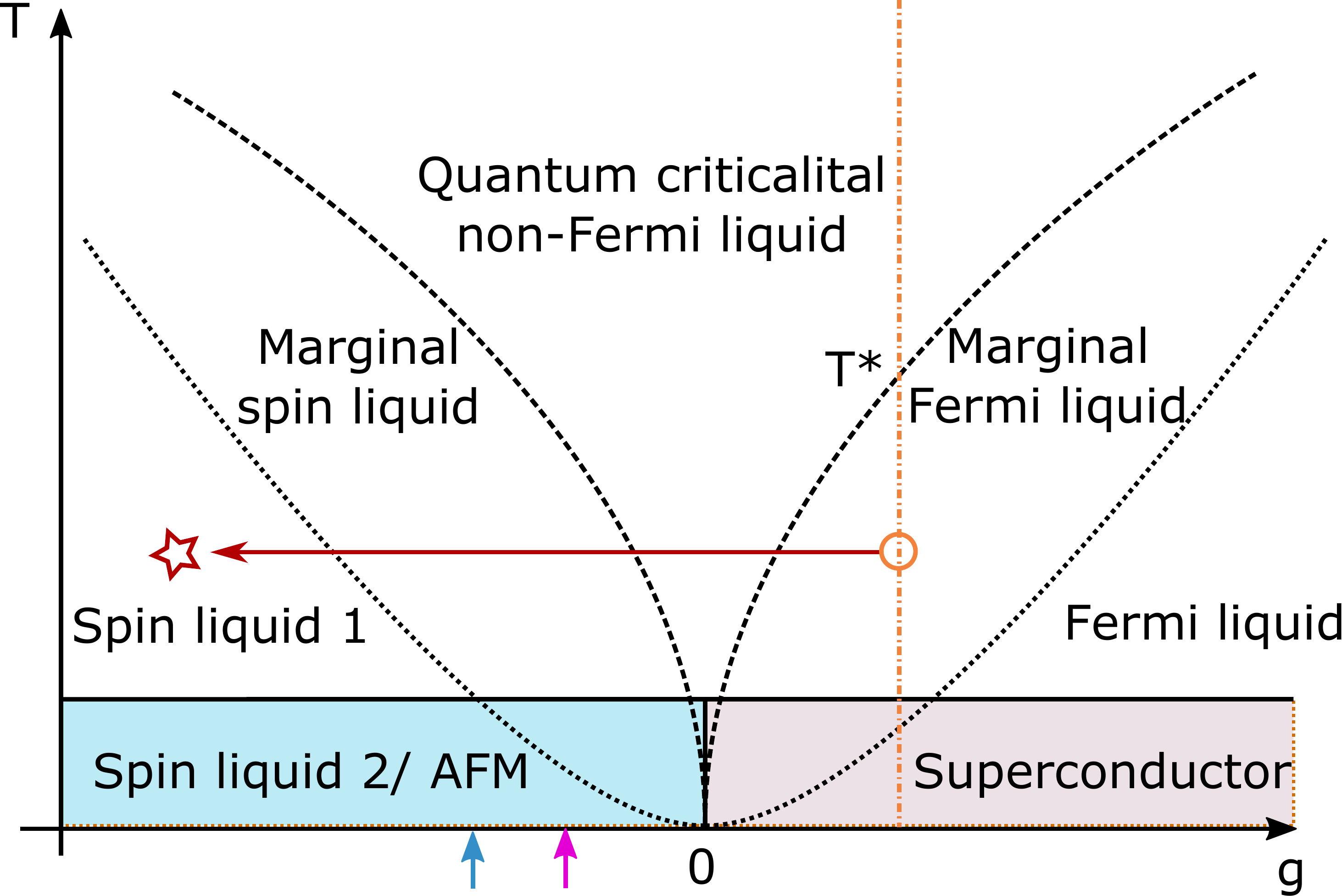}
\caption{Schematic phase diagram (adapted from Ref.~\cite{PhysRevB.78.045109}) showing a would-be continuous Mott transition. Spin liquid 1 represents the spinon Fermi surface state which is unstable at low temperatures. The black dashed lines represent two crossover scales discussed in Ref.~\cite{PhysRevB.78.045109}. The yellow dotted line and the yellow circle represents the estimated position of $\ETBr$. In our interpretation, the pump transiently moves the sample to the spin liquid side as indicated by the red arrow and the red star. The blue arrow and the pink arrow represent the estimated positions of $\ETCl$ and $\ETSL$ respectively.}
\label{Fig:phaseDiagram}
\end{center}
\end{figure}

Furthermore, if the material is metallic but close enough to the Mott transition, we find that the same bosonic excitations that give the superconducting-like response can generate an insulating gap, moving the material into the Mott insulator/spin liquid phase (Fig.~\ref{Fig:phaseDiagram}). We propose that  this could be the case of $\ETBr$. In this interpretation, the photo-induced `superconducting gap' is actually a photo-induced insulating gap! We propose a scaling relation between the extrapolated `superfluid density' and the gap. 

In the following we shall first briefly introduce the slave boson formalism, then discuss the dynamics of a spin liquid after the pump, and finally go back to the current experiment on $\ETBr$.

In the slave-boson formalism~\cite{PhysRevLett.95.036403}, electrons are described by the combination of a bosonic chargon $b (r)$ carrying charge $e$ and a fermionic spinon $f_{\sigma}(r)$ ($\sigma = \uparrow, \downarrow$) carrying spin 1/2, coupled together by an emergent U(1) gauge field $a_\mu (r)$ ($\mu = 0,1,2$). The physical electron annihilation operator is written as $c_{\sigma}(r) = b(r)f_{\sigma}(r)$. The effective Lagrangian is 
$\mathcal{L} =\mathcal{L}_b +\mathcal{L}_f $ where 
\begin{align}
\mathcal{L}_b =  
& \frac{1}{2}|(\partial_t - ia_0)b|^2 - V_b, \nonumber\\
&V_b=\frac{v_b^2}{2}|(\nabla - i\vec{a})b|^2 - \frac{g}{2}|b|^2 + \frac{V}{8}|b|^4 \\
\mathcal{L}_f = & \sum_\sigma \bar{f}_\sigma (i\partial_t - a_0 +\mu_f + \frac{(\nabla +i\vec{a})^2}{2m})f_\sigma
\end{align}
The complex boson field has a potential $V_b$ which is of the standard Ginzburg-Landau form~\cite{fisher1989boson}. For simplicity we have taken a quadratic dispersion for the spinon. The chargon and the spinon are coupled to an internal gauge field $a_\mu$, whose integration eliminates redundant degrees of freedom induced by splitting the electron operator. Integrating out high-energy chargons and spinons generates the dynamics of the gauge field.

Depending on the specific parameters, the slave-boson Lagrangian can describe conventional Fermi liquid phases, exotic spin liquid phases, and the transitions among different phases~\cite{PhysRevB.70.035114,PhysRevB.78.045109,PhysRevLett.95.036403}. In this paper, we focus on cases where the spinon has a Fermi surface. For $g>0$, the boson condenses, $\<b\>\neq 0, c_\sigma\sim\<b\>f_\sigma$, the physical electron operator overlaps with the spinon operator and we recover a Fermi liquid.  
For $g<0$, the boson has a gap which is interpreted as the physical charge gap for doublon and holon excitations. The low energy state is a spin liquid phase with a spinon Fermi surface (spin liquid 1 in Fig.~\ref{Fig:phaseDiagram}). A more detailed study~\cite{PhysRevB.78.045109} shows that there is a continuous transition between the spin liquid and the Fermi liquid, with two crossover scales at finite temperatures (dashed lines in Fig.~\ref{Fig:phaseDiagram}). At low temperatures, this continuous transition can be interrupted by instabilities of the Fermi surface, resulting in superconducting order on the Fermi liquid side and other spin liquid phases or AFM order on the insulating side. We think this theoretical framework  captures well the behavior of the $\kappa\text{-ET}$ family at intermediate temperature/frequency scales and the diversity of phases at low temperatures.

We recall that in the slave boson theory, the physical resistivity is the sum of the boson resistivity and the spinon resistivity, known as the Ioffe-Larkin rule~\cite{PhysRevB.39.8988}.

\begin{equation}
\sigma^{-1} = \sigma^{-1}_b +  \sigma^{-1}_{f}
\label{Eq:ioffe}
\end{equation}
Physically, the gauge theory formulation requires that the boson current and spinon current to be equal to the physical electrical current; the physical voltage induces a gauge voltage that drives the spinon, and the difference between these two voltages drives the boson. Consequently, the two resistivities appear in series.

Now we discuss the pump-probe experiment. In the experiment~\cite{PhysRevX.10.031028}, the pump is chosen to resonantly excite a vibrational mode of the ET molecule. This phonon mode lasts much longer than the pump itself, typically in the order of picoseconds; it serves as the source of periodic drive to the electronic degrees of freedom. An \textit{ab initio} calculation~\cite{PhysRevX.10.031028} found that such an oscillation gives a periodic modulation of $U/t$ of the low energy Hubbard model. In the slave boson theory, this modulation gives an oscillation to the boson mass term. Furthermore, in the presence of a spinon Fermi surface, the emergent gauge field is damped, hence decoupled from the boson field~\cite{PhysRevB.78.045109}. Thus the problem of solving the dynamics under periodic drive reduces to the following Lagrangian for a driven boson.

\begin{equation}
\mathcal{L}_d=\frac{1}{2}|\partial_t b|^2 -  \frac{v_b^2}{2}|\nabla b|^2 + \frac{g}{2}|b|^2 - \frac{V}{8}|b|^4-\lambda\cos(\Omega t)|b|^2.
\end{equation}

\begin{figure}[htb]
\begin{center}
\includegraphics[width=0.3\textwidth]{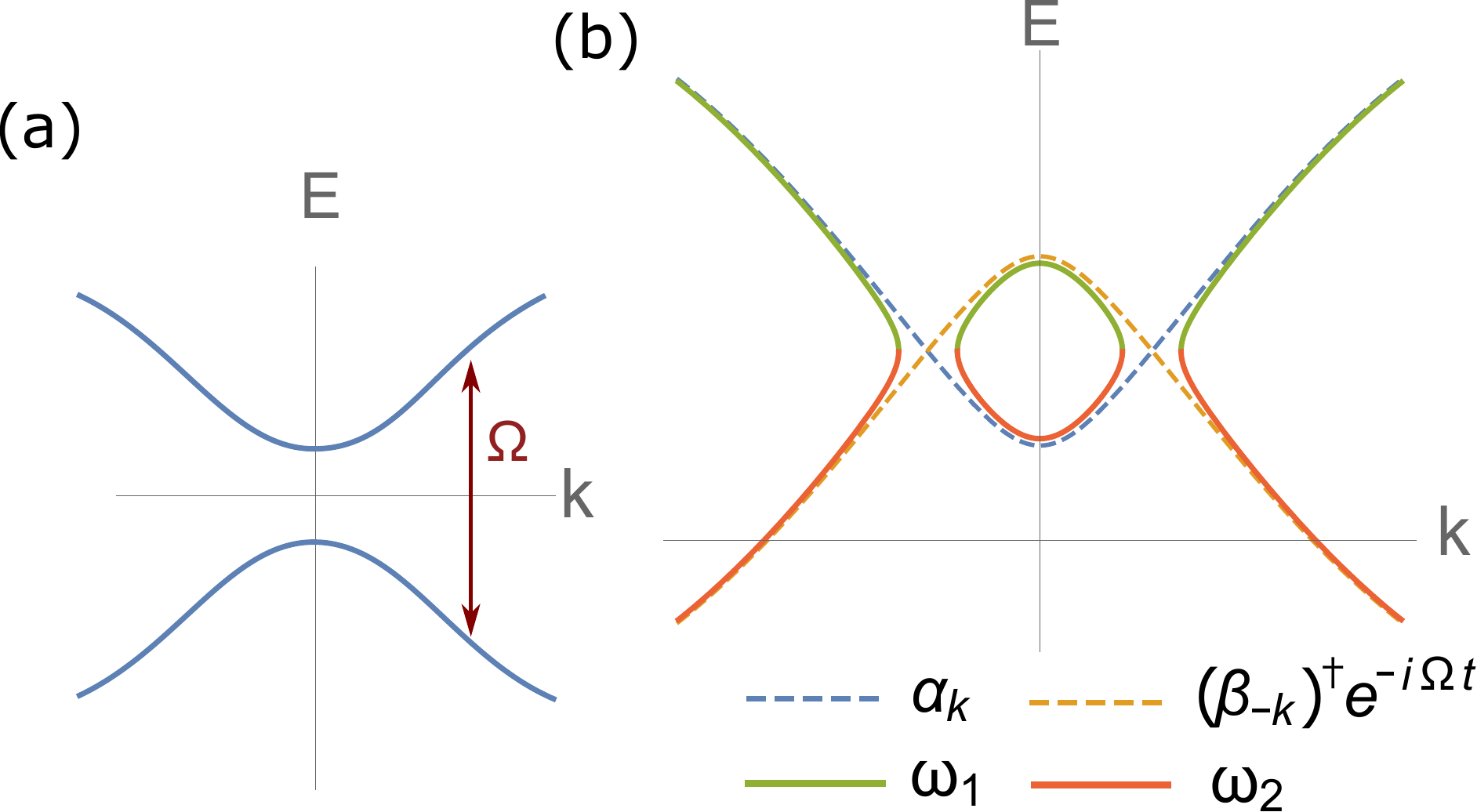}
\caption{(a) Sketch of the gapped bosonic particle and hole band (shown as negative energy for convenience) in a spin liquid. The periodic drive at frequency $\Omega$ resonantly generate particle-hole pairs. (b) Dispersion of a driven boson. The blue/orange dotted line represents the equilibrium dispersion $E_k$/$\Omega-E_k$. The solid lines represent the dispersion under periodic drive. Near the crossings at $\Omega/2$, there is a region (forming a ring in the 2D B.Z.) where the eigen-frequencies are complex and the boson numbers grow exponentially.}
\label{Fig:drivenBosonDispersion}
\end{center}
\end{figure}

For the spin liquid phase, the boson has a positive mass term ($g<0$). The model is the same as the one studied in Ref.~\cite{dai2021photoinduced}, except that now the boson is a fractionalized charge e excitation. For a free boson, the time evolution of the Heisenberg operators can be solved analytically. By standard canonical quantization, we write the boson operator at momentum $k$ as $b_k = \frac{1}{\sqrt{E_k}}(\alpha_k +\beta_{-k}^\dagger)$, where $E_k \simeq \sqrt{(v_b k)^2 + |g|}$. $\alpha_k$ and $\beta_k$ represent particle and anti-particle excitations with a gap $\Delta=\sqrt{|g|}$ which we interpret as the gap for doublon and holon excitations. It is convenient to show the antiparticle dispersion in the lower-half plane as shown in Fig.~\ref{Fig:drivenBosonDispersion}(a). The periodic drive contains the term $e^{-i\Omega t}\a_k^\dagger\b_{-k}^\dagger$ which resonantly excite particle and antiparticle excitations of the boson, visualized as a vertical transition in Fig.~\ref{Fig:drivenBosonDispersion}(a). 
Let us define $\tilde{\b}_k = e^{i\Omega t}\b_k$. For $E_k$ close to $\Omega/2$, we make the rotating wave approximation to obtain

\begin{equation}
    d\left (\begin{array}{c}
\a_k(t)\\
\tilde{\b}_{-k}^\dagger(t)
\end{array}\right)/dt
\simeq -i\left (\begin{array}{cc}
E_k & \frac{\lambda}{2E_k}\\
-\frac{\lambda}{2E_k} & \Omega-E_k\\
\end{array}\right)
\left (\begin{array}{c}
\a_k(t)\\
\tilde{\b}_{-k}^\dagger(t)
\end{array}\right)
\label{Eq:motion}
\end{equation}

The eigenvalues for the matrix are $\omega = \Omega/2 \pm \sqrt{(E_k-\Omega/2)^2 - (\lambda/2E_k)^2}$. When $|E_k -\Omega/2|\lesssim \lambda/\Omega$, the eigenvalues are complex, and the corresponding boson fields grow exponentially (Fig.~\ref{Fig:drivenBosonDispersion}). We note the key difference between driving fermions and bosons lies in the sign of the lower left off diagonal term in Eq. \ref{Eq:motion}. For fermions the sign is positive and the time evolution is unitary, resulting in the well-known Floquet band picture. For bosons the occupation of a ring of $k$ states can grow exponentially.

If we turn on the periodic drive at time zero, we find

\be
\left (\begin{array}{c}
\a_k(t)\\
\tilde{\b}_{-k}^\dagger(t)
\end{array}\right)
\simeq e^{-i\frac{\Omega}{2}t}\left (\begin{array}{cc}
u_k(t) & v_k^*(t)\\
v_k(t) & u_k^*(t)\\
\end{array}\right)
\left (\begin{array}{c}
\a_k(0)\\
\b_{-k}^\dagger(0)
\end{array}\right).
\label{Eq:atbdaggert}
\ee
where
$u_k(t) = \cosh(\theta_k t) + i\frac{\frac{\Omega}{2}-E_k}{\theta_k}\sinh(\theta_k t)$,
$v_k(t) = i\frac{\lambda}{2E_k\theta_k}\sinh(\theta_k t)$ and $\theta_k=\sqrt{(\lambda/2E_k)^2-(E_k-\Omega/2)^2}$.

In Schrodinger's picture, the state at time $t$ is
\begin{align}
|t\>\propto e^{\sum_k\frac{v_k^*(t)}{u_k^*(t)}e^{-i\Omega t}\a_k^\dagger \b_{-k}^{\dagger}}|0\>
\end{align}
which describes a coherent condensate of a pair of particle and hole, reminiscent of an exciton condensate.  However, the binding is very loose and in Ref.~\cite{dai2021photoinduced}, we showed that this non-equilibrium condensate  has a conductivity $i\frac{e^2}{\hbar}\rho_b(t)/\omega$, as long as the effective decay rate of the boson due to scattering and dissipation is smaller than the growth rate $\lambda/\Omega$. The transient `superfluid density' $\rho_b(t)$ is proportional to the density of charge excitations.
Since the pump contains limited energy, $\rho_b(t)$ eventually saturates and decays to zero. Alternatively, while the occupation of the ring in $k$ space can exceed unity, the total density of the holon and the doublon are limited and $\rho_b(t)$ will saturate. This appears to be what is happening in the experiment.

The boson conductivity after the pump should be the sum of the conductivity of the excited bosons and the conductivity of the quasi-static gapped boson, $\sigma_b^\text{q.s.}$, which is present without the pump.
\begin{equation}
    \sigma_b = i\frac{e^2}{\hbar}\frac{\rho_b}{\omega} + \sigma_b^\text{q.s.}.
\end{equation}
By Kubo formula
\begin{align}
\sigma_b^\text{q.s.}=& \frac{i}{\omega}\sum_n|\<n|j_x|0\>|^2[\frac{2E_n}{(\omega+i0^{+})^2-E_n^2}+\frac{2}{E_n}]\nonumber\\
=&\frac{e^2}{16\hbar}(1-(2\Delta/\omega)^2)\theta(\omega-2\Delta)\nonumber\\
&+ i\frac{e^2}{16\pi\hbar}[(2\Delta/\omega)^2 - 1)\ln|\frac{2\Delta+\omega}{2\Delta-\omega}|-4\Delta/\omega],
\label{Eq:sigmaqs}
\end{align}
where $j_x =\sum_k e^2v_b^2k_x|b_k|^2$ is the current operator of the boson.\footnote{We have regularized the response function by subtracting a constant from $\omega\sigma_b^\text{q.s.}$ and demanding that $\omega\sigma_b^\text{q.s.}\rightarrow 0$ when $\omega\rightarrow 0$ as required for non-superconducting states.}
The second line of this equation is for free bosons in 2D. Note that the real part of the conductance shows a gap of $2\Delta$ as expected and saturates to a value of $\frac{e^2}{16\hbar}$.

To find the physical conductivity we use the Ioffe-Larkin rule (Eq.~\ref{Eq:ioffe}). We represent the fermion conductivity by $\sigma_f = \frac{e^2}{\hbar}\frac{\rho_f}{\tau_f^{-1}-i\omega}$, where $\rho_f = n_f/m_f$ is is the Drude weight, $\tau_f^{-1}\ll \rho_f$ comes  from scattering with the gauge fluctuations or from disorder~\cite{PhysRevB.46.5621,lee2020low}.
Since $\rho_b$ is limited by the total energy of the pump, we expect that $\rho_b\ll \rho_f$. Thus, except at very low frequencies and very high frequencies ($\omega\sim\rho_f$), we have $|\sigma_b|\ll|\sigma_f|$; by Ioffe-Larkin rule (Eq.~\ref{Eq:ioffe}), the physical conductivity is approximately the boson conductivity. At very low frequencies we can ignore the conductivity of the quasi-static boson, hence $\sigma^{-1}\simeq \frac{\hbar}{e^2}\frac{-i\omega}{\rho_b} + \frac{\hbar}{e^2}\frac{-i\omega + \tau_f^{-1}}{\rho_f}\simeq \frac{\hbar}{e^2} (\frac{-i\omega}{\rho_b} +\frac{\tau_f^{-1}}{\rho_f})$. Combining this with the quasi-equilibrium boson conductivity, for $\omega\ll\rho_f$, we have
\begin{equation}
\sigma \simeq \frac{e^2}{\hbar}\frac{\rho_b}{-i\omega + (1/\tau_f)\rho_b/\rho_f}+\sigma^\text{q.s.}_b
\end{equation}

The superconducting-like boson conductivity produces a Drude peak at low frequencies. This is different from our previous proposal for cuprates, where the boson is a local excitation and the boson conductivity directly adds to the physical conductivity. However, for $\rho_b\ll \rho_f$, the width of the Drude peak is greatly reduced from the scattering rate of the spinon, and is likely below the current experimental resolution. Nevertheless, it is interesting that by driving the system we can in principle gain access to information on the spinon scattering rate.

For frequencies close to $2\Delta$, the real part of $\sigma^{\text{q.s.}}_b$  grows linearly from zero (Fig.~\ref{Fig:conductivity}(d)). The imaginary part of $\sigma^{\text{q.s.}}_b$ has a logarithmic singularity at $\omega=2\Delta$, which can be seen as a dip in equilibrium conductivity and a small kink in transient conductivity (Fig.~\ref{Fig:conductivity}(c)). When the frequency is further increased, the fermion resistivity dominates over the boson resistivity for $\omega \gtrsim \rho_f$, and the real part of the conductivity starts to decrease (Fig.~\ref{Fig:conductivity}(d)). The imaginary part of the conductivity first rises slightly and then decays as $\rho_f/\omega$ (Fig.~\ref{Fig:conductivity}(c)). To summarize, the conductivity is not much affected by the drive above the Mott gap, but shows "superconducting-like" behavior below the gap: the real part shows a narrow Drude peak and the imaginary part is approximately $\rho_b/\omega$.

Now we discuss driving a material that sits on the metallic side of the Mott transition, which is the case of the experiment on $\ETBr$. In the language of  slave boson theory, the boson is condensed at equilibrium. However, since the metal is close to the Mott transition ($\T*\ll\Omega/2$), the equilibrium condensate has a small superfluid density, which should not affect excitations at the energy scale of the pump. Therefore, we expect the picture of resonantly excited bosonic particle-hole pair to hold qualitatively even in the metallic materials. On the other hand, the considerable bosonic excitations in the resonant region can have a sizable effect on the effective boson mass: due to the interaction $V|b|^4/8$, the excited boson density gives a term $V\<|b(t)|^2\>|b|^2/2$ which can change the sign of $g$ and transiently drive the boson from the superfluid phase into the insulating phase. In other words, the drive can serve as the tuning parameter of the metal-insulator transition, like (reducing) the pressure and the chemical pressure.

Now we analyze the spinon and boson conductivity in $\ETBr$. 
We first consider the equilibrium case. In the quantum critical regime (Fig.~\ref{Fig:phaseDiagram}), the boson conductivity is predicted to be  $\sigma_b = R(\omega, \text{T})e^2/\hbar$, where $R(\omega, \text{T})$ is a number of order unity which weakly depends on temperature and frequency. Thus, in the quantum critical regime, for $\omega\ll\rho_f$, we always have $|\sigma_f|\gg |\sigma_b|$; according to Ioffe-Larkin rule, $\sigma\simeq\sigma_b=R(\omega, \text{T})e^2/\hbar$. 
For $\T*\ll \text{T}\ll t,U$ and $\omega>2\text{T}$, we expected the conductivity to approach a value not far from the free boson conductivity~\cite{PhysRevB.86.245102}, which is given by Eq.~\ref{Eq:sigmaqs} to be $\frac{2\pi}{16}\frac{e^2}{h}$.
Experimentally, at 80K, the height of the conductivity plateau is about $100-200\ \Omega^{-1}\text{cm}^{-1}$~\cite{PhysRevX.10.031028}. With an interlayer space $\sim 15$ angstrom, this conductivity corresponds to $\sim 0.3-0.6 e^2/h$ per ET layer. This is pretty close to the theoretical value and lends confidence to our model. 

In the presence of the drive, at the mean-field level the induced insulating gap is $\Delta(t) = \sqrt{V\<|b(t)|^2\>-g}$. Close to the quantum critical point of the continuous Mott transition, this relation is modified to $\Delta(t) \propto [V\<|b(t)|^2\>-g]^{\nu}$, where $\nu\simeq 0.67$ is the correlation exponent of the 3D XY model~\cite{PhysRevB.78.045109}. Since the excited boson amplitude is proportional to the `superfluid density' $\rho_b$~\cite{dai2021photoinduced}, we propose the following scaling relation between the charge gap and the `superfluid density'
\begin{equation}
\Delta \propto |\rho_b-\rho_0|^{\nu}, 
\end{equation}
where the unknown constant $\rho_0$ can be determined experimentally from where the gap vanishes but the `superfluid density' is nonzero. 
This relation can be tested by varying the pump fluence and the time delays after the pump and at different temperatures. The transient nature of the gap causes residue oscillations of the boson amplitude, which gives further enhancement of the probe light (supplemental material). In our interpretation, this gap is an insulating gap instead of a superconducting gap. The pump actually drives the metal into a non-equilibrium bosonic condensate superimposed with a quasi-static spin liquid phase!

In summary, we propose that driving a system just on the insulating side of the Mott transition may exhibit ``superconducting-like" signals in the conductivity. Further, this mechanism may explain the experiments done on systems just on the metallic/superconducting side because the drive can move the system transiently across the metal-insulator transition. So far, there is no published pump-probe data on spin liquid materials, and it will be very interesting to perform such experiments. In addition to $\ETSL$ , the other spin liquid material  $\beta'\text{-EtMe}_3\text{Sb[Pd(dmit}_2)]_2$ is a particularly attractive candidate. It shows a clear optical gap at ~$600 \text{cm}^{-1}$ \cite{pustogow2018low} which is below the typical resonant phonon energy at $1250 \text{cm}^{-1}$, but not too far below so that the excitation energy of the bosons are at an intermediate energy scale where the fractionalization picture may apply.

Finally, we mention that our general theory of pumped bosons near the Mott transition is directly applicable to cold bosonic atoms trapped in optical lattices. It is known that at integer fillings, the system is described by a relativistic boson theory~\cite{fisher1989boson}. 

Acknowledgement: We thank Andrea Cavalleri, Eugene Demler and Ludwig Mathey for helpful discussions. This research is funded in part by the Gordon and Betty Moore Foundation. PAL acknowledges the support by DOE office of Basic Sciences Grant No. DE-FG02-03ER46076.
\bibliographystyle{apsrev4-2}
\bibliography{photoSC.bib}

\end{document}